# On the alternative description
# of complex holomorphic and Lorentz geometries
# in four dimensions

S.I.Tertychniy *

Institute for Theoretical Physics
University
D-50923 Cologne
Germany

## Abstract

The equivalence of a conformal metric on 4-dimensional space-time and a local field of 3-dimensional subspaces of the space of 2-forms over space-time is discussed and the basic notion of transection is introduced. Corresponding relation is spread to the metric case in terms of notion of normalized ordered oriented transection field. As a result, one obtains a possibility to handle the metric geometry without any references to the metric tensor itself on a distinct base which nevertheless contains all the information on metricity. Moreover, the notion of space-time curvature is provided with its natural counterpart in the transection 'language' in a form of curvature endomorphism as well. To globalize the local constructions introduced, a certain fiber bundle is defined whose sections are equivalent to normalized ordered oriented transection fields and locally to the metric tensor on space-time. The criterion distinguishing the Lorentz geometry is discussed. The resulting alternative method of the description of space-time metricity, dealing with exterior forms foliation alone, seems to be of a power compatible with one of the standard concept based on the metric tensor.

## 0. Introduction

Self and anti-self-dual 2-forms are certainly a very useful tool in a number of fields of mathematical formalism of general relativity and gauge theories. It is well known in particular that the expansion of curvature forms with respect to special self-dual and anti-self-dual bases is one of the the most effective ways to obtain the 'spinor components' of curvature which in turn provide most convenient representation of latter for construction of Einstein equations, classification of conformal curvature and Ricci spinor *etc.* [1,2].

---

* On leave of absence from: The National Research Institute For Physical Technical and Radio-Technical Measurements (VNIIFTRI), Mendeleevo, Moscow Region, 141570, Russia

The electromagnetic field equations are also effectively formulated in terms of self-dual and anti-self-dual forms. That is in particular a reason why self-dual and anti-self-dual 2-forms are the usual working tool (which is not sometimes mentioned explicitly) in a search for solution of Einstein equations especially in vacuum and electrovac cases (see for example [2-7] ). The making use of special families of self-dual and anti-self-dual 2-forms is the base of so-called vector formalism [8,1]. In [9] they are used for representation of Einstein equations in the form of 'closed differential ideal' and in [10] for investigation of integrability of Bianchi identities in vacuum case of Petrov type I. A natural field of applications of mentioned families of 2-forms is the theory of complex Riemannian spaces [11] and especially semi-flat (heavens) or semi-algebraic degenerate (hyperheavens) vacuum solutions of Einstein equations [12-15]. Latter are connected in particular with the notion of non-linear graviton [16,17,19] and the twistor theory [17-23]. Another important developing field where self-dual and anti-self-dual forms serve as indispensable tool are the recent modifications of Ashtekar's approach to canonical gravity formalism [24-34] including applications in instanton theory [29,35-37].

It is important to observe, however, that these advantages, in several cases maybe of more technical character, which are peculiar to notion of self-dual and anti-self-dual 2-forms seem to have a common root of a fundamental nature. It is implied by the fact that the metric uniquely determines and in turn is uniquely determined (up to conformal factor if no normalizing is taken into account) by *the field of linear subspaces* of complex valued 2-forms space $\Lambda^2$ rather than their bases consisting of self-dual or anti-self-dual 2-forms or any other such sort individual 2-forms.

This important fact might be considered in principle as known one (for example, one can found a similar statement in the recent preprint [34]) but it has more or less a status of a 'folk theorem' and probably has not been carefully investigated yet. The analysis yields however the following additional results. The conformal metric class is connected with the field of 3-dimensional subspaces of $\Lambda^2$ and 'almost every' such field determines some conformal metric class. The 'field of subspaces' associated with *some* metric can be characterized by certain simple 'intrinsic' algebraic properties expressed in terms of binary exterior product alone. In order to describe a *metric* (rather than its conformal class) some normalizing procedure including consideration of additional relations of *ordering* and *orientation* (they are properly introduced and discussed below) has to be introduced. Furthermore such an important characteristic of metric spaces as the curvature can be equivalently represented in this language of 'fields of subspaces' as well.

It is essential that all the above constructions does not require introduction of metric tensor. The relevant 'arena' is here the foliation of exterior forms alone.

These facts suggest possibility to build up a closed realization of 4-dimensional Riemannian structure in a way independent with orthodox form of Riemannian geometry based on notion of metric tensor. It is ousted here by the notion of *transection* of the complex 2-forms space $\Lambda^2$ which means essentially decomposition of $\Lambda^2$ into a direct sum of two complementary subspaces, the *lobes* of transection, obeying simple algebraic relations.

It seems difficult to estimate *a priory* all possible advantages (maybe disadvantages in some respects) of such an alternative description of metric geometry but one may still hope that it might give rise to a better insight into the gravitation theory.

There is a number of works intersecting with subject of this paper [31,38-44]. Maybe the most closed and profound work which is to be referred to still remains [38] (see also [39]). Here relations between the metric and special bases of self- and anti-self-dual 2-forms ($S$-forms in our terminology) are investigated and a number of statements of our section 1 have their counterparts here. It is mentioned also that the metric tensor implies the 'wedge orthogonal' decomposition of the 2-forms space into 2-dimensional subspaces but inverse implication is not discussed. There is also a number of works where the metric is deduced from considerations of self- and anti-self-dual 2-forms but particular bases rather than subspaces that they span serve as main object of consideration [30,33,40,41,46]. In the work [31] the crucial role of the eigenspaces of dualizing operator is briefly mentioned but the distinct from our tools (totally null $\alpha$- and $\beta$-triplanes) are exploited. The connection of metric with dual operator is also studied in [42-44].

We intend to give here one of the possible closed expositions of the method of 4-dimensional metric geometry description in the language of exterior forms foliations. In section 1 the basic definition of transection is given, the canonical bases ($S$-bases) of 2-forms space are defined together with corresponding invariance (redundancy) group and the relations of these objects with conformal metric are discussed. The case of metric geometry is considered in section 2 ant the basic notion of normalized ordered oriented transection is introduced. In section 3 the closed 'intrinsic' description of transection lobe is obtained which allows to introduce alternative definition of transection. The relation of $\Lambda^2$ subspaces associated with conformal metric and the Grassmann manifold $\mathbb{C}G_{6,3}$ is established. In section 4 the connection of field of normalized ordered oriented transections and certain fiber bundles (transection bundles) is discussed. In section 5 the representation of Riemannian curvature in transection language is investigated. It is analyzed in section 6 what features of the metric description in terms of transections corresponds to the Lorentz metric case. In section 7 a possible way of real Lorentz metric representation without complexification of 2-forms space is outlined. In section 8 the main results are accumulated and briefly exposed. In conclusion two levels of metric geometry description in terms of 2-forms are discussed. Appendix (supmitted separately) contains the proofs that are omitted in main text.

## 1. Transection and its algebraic properties

In the first sections we shall deal with algebraic relations in the Grassmann algebra over a single point unless opposite is stated.

Let $\Lambda$ denote 4-dimensional vector space of 1-forms over the field of complex numbers $\mathbb{C}$. The binary operation of exterior multiplication maps its direct square $\Lambda \times \Lambda$ onto 6-dimensional space $\Lambda^2 = \Lambda \wedge \Lambda$ of 2-forms over $\mathbb{C}$. In turn, direct square of $\Lambda^2$ is mapped by exterior multiplication onto 1-dimensional space $\Lambda^4$ of 4-forms.

⟨ **Definition 1.1** The *transection* of the space $\Lambda^2$ of 2-forms over $\mathbb{C}$ is its decomposition into a direct sum of two 3-dimensional subspaces $^+\Lambda^2$ and $^-\Lambda^2$ such that any two elements chosen from different subspaces are wedge-orthogonal.

In other words $\Lambda^2$ is to be represented as follows

$$\Lambda^2 = {}^+\Lambda^2 \oplus {}^-\Lambda^2, \quad \dim {}^+\Lambda^2 = \dim {}^-\Lambda^2 \ (= 3), \quad {}^+\Lambda^2 \wedge {}^-\Lambda^2 = 0.$$

The subspaces $^{\pm}\Lambda^2$ will be called *the lobes* of transection or simply lobes below. They enter the transection definition in a completely symmetric fashion and thus have to be jointed to a *disordered* pair. It will be convenient, however, to impose some order to the two element sequence of lobes and call the first of them, say, $^{+}\Lambda^2$ *undotted* (or *right, heavenly*) lobe while the second one, $^{-}\Lambda^2$, *dotted* (respectively *left, hellish*) lobe. (These terms receive some justification in the notations that will be introduced below and are in common use). Since the ordering of lobes is an additional structure introduced for convenience all the facts that concern the transection itself are to be invariant with respect to permutation $^{+}\Lambda^2 \Longleftrightarrow {}^{-}\Lambda^2$ which occasionally will be called a reflection of lobes below.

The wedge-orthogonality of lobes implies some specific properties of the above decomposition $\Lambda^2 = {}^{+}\Lambda^2 \oplus {}^{-}\Lambda^2$.

↯ **Lemma 1.2** [1)]    Let $\alpha$ be an arbitrary nonzero 2-form. if $\alpha \wedge \beta = 0$ for every $\beta \in {}^{\pm}\Lambda^2$ (equivalent record: $\alpha \wedge {}^{\pm}\Lambda^2 = 0$) then $\alpha \in {}^{\mp}\Lambda^2$.

↯ **Corollary 1.3**    For every nonzero element of a lobe there exist the element of the same lobe which is not wedge-orthogonal to the first one.

One sees that specifying a lobe of transection, the opposite lobe is unambiguously specified as well. Indeed, if the triad of 2-forms $\alpha_j, \; j = 1, 2, 3$ constitute the basis of the first lobe the second one spans just by the forms $\beta$ obeying equations $\beta \wedge \alpha_j = 0$. We formulate this fact in the form of

↯ **Proposition 1.4**    If a 3-dimensional subspace of $\Lambda^2$ is a lobe of some transection this transection is specified by it in a unique way (together with the second lobe).

In accordance with transection definition the lobe is characterized by its wedge-orthogonality with respect to the second one which is defined just in the same way in terms of relation to the first. It seems reasonable to break this definition circle and to try to describe a lobe in terms of its intrinsic algebraic properties. This will be realized in the section 3 where some corresponding criteria are stated and here we continue with a description of the algebraic structure of transection and its lobes.

↯ **Lemma 1.5**    Every lobe of transection contains a nonzero simple 2-form.

**Remark 1.6**    This lemma could be considered as particular case of a slightly more general statement: Every more than 1-dimensional linear subspace of the space of 2-forms over $\mathbb{C}$ contains a nonzero simple element. (For real forms this is not the case).

A similar property concerns with non-simple 2-forms as well in according with

↯ **Lemma 1.7**    Every lobe of transection contain a non-simple 2-form.

The next statement describes a content of a lobe in more details.

↯ **Lemma 1.8**    Every lobe contains a triad of elements $\alpha, \beta, \gamma$ such that

$$\alpha \wedge \alpha = 0 = \beta \wedge \beta = \alpha \wedge \gamma = \beta \wedge \gamma, \quad \gamma \wedge \gamma \neq 0 \neq \alpha \wedge \beta. \tag{1.1}$$

---

[1)] All the necessary proofs can be found in Appendix which is contained in the second part of this paper submitted separately.

**Remark 1.9** All these 2-forms are obviously nonzero and linearly independent in total. They constitute a basis of the lobe.

**Remark 1.10** One may assume without loss of generality a fulfillment of the additional equation

$$\alpha \wedge \beta = -2\gamma \wedge \gamma. \tag{1.2}$$

These properties of the lobes imply the following important representation of a transection in terms of the special bases of its lobes:

ι **Theorem 1.11** For every transection there exists a basis $\theta_{A\dot{B}}$ of the space $\Lambda$ of 1-forms such that the collection of 2-forms consisting of

$$S_{AB} = S_{BA} = \tfrac{1}{2} \epsilon^{\dot{K}\dot{L}} \theta_{A\dot{K}} \wedge \theta_{B\dot{L}} \tag{1.3a}$$

constitutes the basis of the lobe $^+\Lambda^2$ and the collection

$$S_{\dot{A}\dot{B}} = S_{\dot{B}\dot{A}} = \tfrac{1}{2} \epsilon^{KL} \theta_{K\dot{A}} \wedge \theta_{L\dot{B}} \tag{1.3b}$$

constitutes the basis of $^-\Lambda^2$.

Here the dotted and undotted indices run over the (distinguished) sets $\{0,1\}$ and $\{\dot{0},\dot{1}\}$ respectively.

The notations used here are borrowed from the spinor algebra and some other usual modes peculiar to this field will be used in what follows without additional elucidation. As a general reference we can point out to [45] although the indices manipulation rule adopted here is $\iota^B = \iota_A \epsilon^{AB}$. Such definition is used in [15] and it differs from that in [45].

We shall also employ so called *summed spinor indices* parallely with ordinary ones for objects symmetric in indices of the same kind. Every component of the object possessing $m$ dotted and $n$ undotted spinor indices and symmetric with respect to them can be identified by *the pair* of indices (or unique one if only one sort indices presents), the first of them running over the set $0, 1, \ldots, 2^{m-1}$ and the second over the set $\dot{0}, \dot{1}, \ldots, \dot{2}^{n-1}$ respectively (of course, the cases $m = 0$ or $n = 0$ must be treated separately in the obvious way).

ι **Corollary 1.12** For every transection there exist the basis $S_{AB}$ of undotted lobe and the basis $S_{\dot{A}\dot{B}}$ of dotted lobe such that

$$S_{AB} \wedge S^{CD} = \tfrac{1}{3} \delta^C_{(A} \delta^D_{B)} \, S_{KL} \wedge S^{KL} \neq 0 \qquad \Rightarrow \qquad S_{(AB} \wedge S_{CD)} = 0; \tag{1.5a}$$

$$S_{\dot{A}\dot{B}} \wedge S^{\dot{C}\dot{D}} = \tfrac{1}{3} \delta^{\dot{C}}_{(\dot{A}} \delta^{\dot{D}}_{\dot{B})} \, S_{\dot{K}\dot{L}} \wedge S^{\dot{K}\dot{L}} \neq 0; \qquad \Rightarrow \qquad S_{(\dot{A}\dot{B}} \wedge S_{\dot{C}\dot{D})} = 0; \tag{1.5b}$$

$$S_{AB} \wedge S_{\dot{A}\dot{B}} = 0; \tag{1.6}$$

$$S_{KL} \wedge S^{KL} + S_{\dot{K}\dot{L}} \wedge S^{\dot{K}\dot{L}} = 0. \tag{1.7}$$

The equations (1.5a) or (1.5b) are merely another form of relations (1.1), (1.2); (1.6) is equivalent to the condition of the wedge-orthogonality of $^+\Lambda^2$ and $^-\Lambda^2$; (1.7) is ensured by the corresponding normalizing of bases. All the relations $(1.5-7)$ are the consequences of the eqs. (1.3).

**Remark 1.13** The basis $S_{AB}$ of the undotted lobe is in fact identical (i.e. differs only notationally) to any basis that obeys the relations $(1.1, 2)$. Since the latter are equivalent to $(1.5a)$ we can conclude that $(1.5a)$ is the necessary and sufficient condition ensuring the basis $S_{AB}$ to be expressed in terms of some basis $\theta_{A\dot{B}}$ of $\Lambda$ by means of the eq. $(1.3a)$. The same is true, of course, for $(1.5b)$ and $(1.3b)$ if one deals with the dotted lobe.

≀ **Definition 1.14** The basis of a lobe with the spinor indexing which obeys the relations $(1.5a)$ (for undotted lobe) or $(1.5b)$ (for dotted one) will be called $S$-*basis*. The elements of $S$-basis will be called $S$-*forms*.

There is no a good name for these objects in the literature. Our one reflects more the notations (introduced initially in [13]) and one might not be gratified by it as well but since the need to refer to '$S$-forms' or '$S$-bases' will meet in almost every paragraph of the text below it is necessary to introduce some abbreviated name even if it would seem to be not very successful.

In accordance with remark 1.13 every $S$-basis can be expressed in terms of some 1-forms basis by means of eqs. $(1.3a)$ or $(1.3b)$ and conversely the possibility of such representation implies the fulfillment of $(1.5)$.

Keeping this in mind let us introduce for convenience the following *auxiliary definitions*. We shall name the $S$-bases of the undotted and dotted lobes *fitted* by some tetrad $\theta_{A\dot{B}}$ if these objects obey the eqs. $(1.3)$. If these equations hold only *up to a factor* common for all the elements of certain $S$-basis from the pair these $S$-bases will be called *conformally fitted*. Finally, the tetrad $\theta_{A\dot{B}}$ will be called *fitting* (with respect to given transection) if the formulae $(1.3)$ define the bases (fitted by definition) of the lobes of this transection. Then the theorem 1.11 can be recast as follows.

≀ **Theorem 1.11'** There exists a fitting tetrad for every transection.

The knowledge of $S$-bases is of course sufficient for the specifying of transection but such its a representation manifests an essential disadvantage to be highly redundant. Our current purpose is to describe the corresponding arbitrariness. It may be realized as a certain Lie group of transformations of $S$-forms.

At first, it is evident that all the $S$-forms can be multiplied by any nonzero number. We shall name such transformations *conformal*. If the eq. $(1.7)$ is observed the conformal factors for the undotted and dotted lobes must coincide or at most differ by sign.

Further, all the properties of undotted $S$-forms are invariant with respect to following 1-complex-parameter groups of transformations:

$$g_0(\rho), \quad \rho \in \mathbb{C}: \qquad (g_0 S)_0 = S_0, \quad (g_0 S)_1 = S_1 + \rho S_0, \tag{1.8a}$$
$$(g_0 S)_2 = S_2 + 2\rho S_1 + \rho^2 S_0,$$

$$g_1(\rho), \quad \rho \in \mathbb{C}\backslash 0: \qquad (g_1 S)_1 = S_1, \quad (g_1 S)_0 = \rho^2 S_0, \quad (g_1 S)_2 = \rho^{-2} S_2; \tag{1.8b}$$

$$g_2(\rho), \quad \rho \in \mathbb{C}: \qquad (g_2 S)_2 = S_2, \quad (g_2 S)_1 = S_1 + \rho S_2, \tag{1.8c}$$
$$(g_2 S)_0 = S_0 + 2\rho S_1 + \rho^2 S_2,$$

and the cyclic second order group generated by transformation

$$g_\uparrow: \qquad (g_\uparrow S)_0 = S_2, \quad (g_\uparrow S)_1 = -S_1, \quad (g_\uparrow S)_2 = S_0. \tag{1.8d}$$

The following commutation rules hold

$$g_1(\rho) \circ g_0(\sigma) = g_0(\rho^2 \sigma) \circ g_1(\rho),$$
$$g_2(\sigma) \circ g_1(\rho) = g_1(\rho) \circ g_2(\rho^2 \sigma),$$
$$g_2(\rho) \circ g_0(\sigma) = g_0(\sigma\Sigma) \circ g_1(\Sigma) \circ g_2(\sigma\Sigma) \quad \text{where } \Sigma = (1 + \rho\sigma)^{-1} \quad \text{if } \rho\sigma \neq -1,$$
$$= g_0(\rho^{-1}) \circ g_1(\rho^{-1}) \circ g_2\left(-\rho^{-1}(1+\rho\sigma)\right) \circ g_\uparrow \qquad \text{if } \rho \neq 0, \qquad (1.9)$$
$$g_\uparrow \circ g_0(\rho) = g_2(-\rho) \circ g_\uparrow,$$
$$g_\uparrow \circ g_2(\rho) = g_0(-\rho) \circ g_\uparrow,$$
$$g_\uparrow \circ g_1(\rho) = g_1(\rho^{-1}) \circ g_\uparrow.$$

They imply that all possible compositions of the transformations (1.8) constitute the Lie group which we denote $G'$. It is isomorphic to $SO(3, \mathbb{C})$. [Indeed, transformations (1.8) are represented with respect to basis $\{(S_0 + S_2), S_0 - S_2, 2S_1\}$ precisely by complex orthogonal matrices; next, $G'$ is obviously 6-real-dimensional and connected; and finally it possesses two-fold covering as we shall see below.] Since the relation to the orthogonal group we shall name the transformations (1.8) *rotations* (and (1.8d), occasionally, *discrete rotation*). Notice that the second and third lines of (1.9) enable one to replace the discrete rotation in favor some composition of 'continuous' ones throughout (it will not have however 'canonical ordered' form $g_0 \circ g_1 \circ g_2$), for example the representation $g_\uparrow = g_2(1) \circ g_0(-1) \circ g_2(1)$ holds. It is useful however to keep $g_\uparrow$ in the list (1.8). It use drastically simplifies a lot of algebraic proofs. One of the possible its application is the constructing of the atlas covering the group $G'$: it consists of two charts, the first of them contains all the rotations of the form $g_0 \circ g_1 \circ g_2$ and the second all the rotations $g_0 \circ g_1 \circ g_2 \circ g_\uparrow$. Eqs. (1.9) imply that this indeed will be the atlas. (We shall omit below the composition sign '$\circ$' in formulae).

Eqs. (1.8) involve only one sort of $S$-forms (undotted ones). There exist exactly the same arbitrariness for a choice of the elements of dotted $S$-basis. To describe it, one may simply add the dots over all the indices in (1.8) (that yields *inversely dotted* with respect to (1.8) equations). We shall consider the corresponding transformation group $\dot{G}'$ to be distinct (although isomorphic) with $G'$. Its elements will be denoted by the same kernel $g$ but with a dot above: $\dot{g}$.

The groups $G'$ and $\dot{G}'$ commute.

≀ **Lemma 1.15** Any two $S$-bases of any chosen lobe of transection are connected by the composition of a conformal transformation (possibly trivial) and a rotation.

The rotations of $S$-basis are naturally connected with the rotations of tetrad. One has the following

≀ **Lemma 1.16** Let two $S$-bases $S_{AB}$ and $\tilde{S}_{AB}$ of the undotted lobe be fitted with the same $S$-basis of the dotted lobe by means of the tetrads $\theta_{A\dot{B}}$ and $\tilde{\theta}_{A\dot{B}}$ respectively. Then apart from the obvious possible conformal multiplication to one of the fourth roots of the unit the tetrads $\theta_{A\dot{B}}, \tilde{\theta}_{A\dot{B}}$ are connected by a certain composition of transformations from the following list:

$$g_0(\rho), \quad \rho \in \mathbb{C}: \qquad (g_0\theta)_{0\dot{A}} = \theta_{0\dot{A}}, \qquad\qquad (g_0\theta)_{1\dot{A}} = \theta_{1\dot{A}} + \rho\theta_{0\dot{A}}; \quad (1.10a)$$

$$g_1(\rho), \quad \rho \in \mathbb{C}\backslash 0: \quad (g_1\theta)_{0\dot{A}} = \rho\theta_{0\dot{A}}, \qquad (g_1\theta)_{1\dot{A}} = \rho^{-1}\theta_{1\dot{A}}; \qquad (1.10b)$$

$$g_2(\rho), \quad \rho \in \mathbb{C}: \qquad (g_2\theta)_{0\dot{A}} = \theta_{0\dot{A}} + \rho\theta_{1\dot{A}}, \quad (g_2\theta)_{1\dot{A}} = \theta_{1\dot{A}}; \qquad (1.10c)$$

$$g_\uparrow: \qquad\qquad\qquad (g_\uparrow\theta)_{0\dot{A}} = \theta_{1\dot{A}}, \qquad\qquad (g_\uparrow\theta)_{1\dot{A}} = -\theta_{0\dot{A}}. \qquad (1.10d)$$

The 'inversely dotted' statement holds as well.

**Remark 1.17** The set of compositions of transformations (1.10) (that also will be called rotations) constitutes the group isomorphic to $SL(2,\mathbb{C})$. The correspondence $(1.10)\Longleftrightarrow(1.8)$ is the two-fold universal covering of $SO(3,\mathbb{C})$ by $SL(2,\mathbb{C})$. We will denote below as $G$ just the $SL(2,\mathbb{C})$ transformations (1.10) and will consider $G'$ as the representation of the $G$. It is worth noting that the discrete rotations (1.10d) constitute the fourth order cyclic group. It coincides in fact with the so-called "Sach's transformations" [11].

Another copy of $SL(2,\mathbb{C}) = \dot{G}$ (dotted rotations) acts on tetrads by means of the *inversely dotted* formulae (1.10) (one may easily define what this precisely means) and has the $SO(3,\mathbb{C})$-representation acting on the dotted $S$-bases and remaining undotted $S$-forms unaffected.

The property to be conformally fitted does not really imply any special incorporation of individual undotted and dotted $S$-bases to a distinguished pair as the following corollary of the lemma 1.15 manifests:

≀ **Corollary 1.18** Any undotted and dotted $S$-bases are conformally fitted.

The prolongation of the property to be fitted from the level of $S$-bases to the level of tetrads is described by the following

≀ **Lemma 1.19** Let the $S$-bases $S_{AB}$ and $S_{\dot{A}\dot{B}}$ be conformally fitted by the tetrad $\theta_{A\dot{B}}$ and the $S$-bases of the same transection $\tilde{S}_{AB}$ and $\tilde{S}_{\dot{A}\dot{B}}$ be conformally fitted by the tetrad $\tilde{\theta}_{A\dot{B}}$ respectively. Then $\theta_{A\dot{B}}$ and $\tilde{\theta}_{A\dot{B}}$ are mutually connected by the composition of (i) conformal transformation, (ii) undotted rotation and (iii) dotted rotation.

Notice that *conformal reflection* of tetrad $\theta_{A\dot{B}} \to -\theta_{A\dot{B}}$ coincides with $g_\uparrow^2$ and is, therefore, a rotation.

As a consequence of the corollary 1.18 and the above lemma one has

≀ **Corollary 1.20** Any two tetrads fitting with respect to the same transection are connected by composition of conformal transformation and rotations of both types

Now we are able to state the first basic result:

≀ **Theorem 1.21** A transection is equivalent to conformal metric.

It should be noted here that we have never mentioned above the complex conjugation operation and thus the metrics here are assumed to be complex valued ones.

It is easy to see the way of practical connection of the conformal metric and transection. A transection equipped every lobe with a family of $S$-bases in accordance with theorem 1.11. Due to the corollary 1.18 any pair of $S$-bases of different lobes is fitted by some tetrad $\theta_{A\dot{B}}$. Having any such tetrad, one may introduce the following symmetric (and obviously non-degenerated) second order tensor $\mathbf{g}$

$$\mathbf{g} = \theta^A{}_{\dot{B}} \otimes \theta_A{}^{\dot{B}} \tag{1.11}$$

It is easy to see that it does not depend on the arbitrariness involved in its definition (apart of the conformal rescaling). It is just this metric tensor which is provided by the theorem above (see also its proof in Appendix for more details).

To 'extract' the transection from the metric, let us notice at first that any non-degenerate symmetric tensor $\mathbf{g}$ can be reduced (over $\mathbb{C}$) to a diagonal form

$$\mathbf{g} = \Theta^a \otimes \Theta^a, \qquad a = 1, 2, 3, 4 \tag{1.12}$$

where 1-forms $\Theta^a$ constitute the basis of $\Lambda$. The corresponding spinor indexing may be introduced by the transition to tetrad $\theta_{A\dot{B}}$ in the following way:

$$\theta_{0\dot{0}} = \Theta^3 + i\Theta^4, \qquad \theta_{1\dot{1}} = -\Theta^3 + i\Theta^4,$$
$$\theta_{0\dot{1}} = \Theta^1 + i\Theta^2, \qquad \theta_{1\dot{0}} = -\Theta^1 + i\Theta^2.$$

Then the eq. (1.11) turns out to be equivalent to (1.12) and the formulae (1.3) yield the $S$-basis of undotted and dotted lobes of some transection. The latter does not depend on the arbitrariness in the choice of tetrad $\Theta^a$ (see the proof in Appendix).

**Corollary 1.22** All the elements of transection lobes are invariant with respect to dualyzing (Hodge) operator determined by metric which in turn is associated with transection.

The lobes of transections consist of anti-self-dual (undotted lobe) and self-dual (dotted lobe) 2-forms and this property of $S$-forms is well known and widely used (see for example [33] and references therein). It is reasonable however to apply possible definition of lobes as (anti-)self-dual subspaces with respect to Hodge operator only if one begins from the metric tensor whose knowledge is necessary to specify Hodge dual. On the contrary, if it is a transection that is considered as primary object the metric itself should be defined as determining Hodge operator with necessary properties. In any case self-duality does not play a noticeable role in our approach.

There is also explicit formula expressing metric tensor (1.11) in terms of holonomic components of $S$-forms. It can be represented as follows (*cf.* [46,30,33]):

$$\sqrt{\det \mathbf{g}}\, \mathbf{g} = \tfrac{4}{3} \varepsilon^{\alpha\beta\gamma\delta} S_K{}^L{}_{\rho\alpha} S_L{}^M{}_{\beta\gamma} S_M{}^K{}_{\delta\sigma}\, dx^\rho \otimes dx^\sigma. \tag{1.13}$$

Here $x^\rho$ are some coordinates, $S_K{}^L{}_{\alpha\beta}$ are the components of $S$-forms with respect to holonomic basis $dx^\rho \wedge dx^\sigma$, *i.e.* $S_{AB} = S_{AB\,\rho\sigma} dx^\rho \wedge dx^\sigma$.

The sense of eq. (1.13) becomes quite transparent if one use 'tetrad' components of $S$-forms instead of holonomic ones. Indeed, if we define tetrad components of $S$-forms $S_K{}^L{}_{M\dot{M}N\dot{N}}$ by means of expansion $S_K{}^L = S_K{}^{LM\dot{M}N\dot{N}} \theta_{M\dot{M}} \wedge \theta_{N\dot{N}}$ then

$$S_K{}^L{}_{M\dot{M}N\dot{N}} = \tfrac{1}{2} \delta_N^L \epsilon_{KM} \epsilon_{\dot{M}\dot{N}}.$$

and straightforward calculation yields

$$\epsilon^{abcd}\,S_K{}^L{}_{ra}\,S_L{}^M{}_{bc}\,S_M{}^K{}_{ds}\,\theta^r\otimes\theta^s =$$
$$\left(\epsilon^{AD}\epsilon^{BC}\epsilon^{\dot A\dot C}\epsilon^{\dot B\dot D}-\epsilon^{AC}\epsilon^{BD}\epsilon^{\dot A\dot D}\epsilon^{\dot B\dot C}\right)S_K{}^L{}_{R\dot R A\dot A}\,S_L{}^M{}_{B\dot B C\dot C}\,S_M{}^K{}_{D\dot D S\dot S}\,\theta^{R\dot R}\otimes\theta^{S\dot S}$$
$$=-\tfrac{3}{4}\,\epsilon_{RS}\epsilon_{\dot R\dot S}\,\theta^{R\dot R}\otimes\theta^{S\dot S}=\tfrac{3}{4}\,\mathbf{g}.$$

On the other hand $\epsilon^{abcd}\,S_K{}^L{}_{ra}\,S_L{}^M{}_{bc}\,S_M{}^K{}_{ds}\,\theta^r\otimes\theta^s = \det\|\theta^\alpha_a\|\,\epsilon^{\alpha\beta\gamma\delta}\,S_K{}^L{}_{\rho\alpha}\,S_L{}^M{}_{\beta\gamma}$ $S_M{}^K{}_{\delta\sigma}\,dx^\rho\otimes dx^\sigma$ where $\theta^\alpha_a$ are defined by equation $\theta^\alpha_a\theta^a = dx^\alpha$. The determinant of matrix $\|\theta^\alpha_a\|$ is determined from relation $\mathbf{g}_{\alpha\beta}\theta^\alpha_a\theta^\beta_b = \mathbf{g}_{ab}$ where $\mathbf{g}_{12} = 1 = \mathbf{g}_{34}$ and other $\mathbf{g}_{ab}$ vanish which yields $\det\|\theta^\alpha_a\| = (\det\mathbf{g})^{-1/2}$. Then the eq. (1.13) follows.

## 2. Normalized ordered oriented transection and metric tensor

The main subject of this work is the description of a metric geometry rather than a conformal one. The reduction of geometry may be achieved by means of a certain normalizing procedure although latter seems to be a bit less natural than that we have seen above. We begin with

≀ **Definition 2.1**  Normalizer is a nonzero 4-form.

Let some normalizer $\omega$ be chosen and fixed. Let some transection be given.

≀ **Definition 2.2**  The $S$-basis $S_{AB}$ of undotted lobe will be called *normalized* if

$$S_{AB}\wedge S^{AB} = 3i\omega. \qquad (2.1a)$$

The dotted $S$-basis $S_{\dot A\dot B}$ is normalized if

$$S_{\dot A\dot B}\wedge S^{\dot A\dot B} = -3i\omega. \qquad (2.1b)$$

(cf. (1.7)).

If $S_{AB}$ is a normalized $S$-basis then $\tilde S_{AB} = -S_{AB}$ is a normalized $S$-basis as well. It is reasonable to consider them to be oppositely oriented. It can be realized in frames of the following

≀ **Definition 2.3**  We shall say that two normalized $S$-bases of the same lobe are of equal orientation if they are connected by a rotation only. The class of all $S$-bases those every pair is of equal orientation will be called the orientation of a lobe.

It is evident that $S$-bases are not of equal orientation if and only if they are connected apart from a rotation by the conformal reflection also, i.e. $\tilde S_{AB} = -gS_{AB}$. The 'equal orientation' is the equivalence relation and there are exactly two classes of normalized $S$-bases of equal orientation and they have no common elements. One may say therefore that any two normalized $S$-bases that are not of equal orientation are of opposite orientation.

All the above remarks on the orientation are of course valid for the dotted $S$-bases as well.

The notion of orientation of lobe enables one to specify the corollary 1.18 to the case of normalized transections.

> **Proposition 2.4**  Any $S$-basis of the chosen orientation specified by a certain lobe of normalized transection is fitted with all $S$-bases of one and only one of two possible orientations of the opposite lobe and is *not* fitted with any $S$-basis of another orientation.

This property means that fixing orientation of one lobe one can distinguish in a canonical way one of two possible orientations of another lobe (that which contains $S$-bases fitted with properly oriented $S$-bases of the first lobe). Thus orientations of opposite lobes are canonically joined in pairs.

> **Definition 2.5**  A pair of canonically co-ordinated orientations (in the way described above) is named *orientation of normalized transection*.

There are exactly two different orientations of normalized transection.

Let us notice now that every undotted $S$-basis can be expressed in terms of some tetrad by the formula (1.3a) and, moreover, such a tetrad is specified up to dotted rotations. The eq. (1.11) then determines a metric which is not affected by arbitrariness in the choice of tetrad. Further, if one changes the orientation of $S$-basis (for example, by means of conformal reflection) then this will result in the change of a metric sign. The second possible arbitrariness affecting the metric is the different *ordering* of the lobes *i.e.* a choice what a lobe will be called undotted one. The undotted lobe is not 'better' in any respects than the dotted one and *vice versa*. Thus, initially, any of two lobes may be called 'undotted'. If the fixed normalizing of transection is observed, change of 'initial ordering' of lobes yields the additional factor $i$ as it concerns with a metric. The formal cause of this factor is the eq. (1.7), one can also see from (2.1) that the same effect is achieved by the reversing of normalizer sign with constant lobes. We can assemble these facts in the form of

> **Proposition 2.6**  The pair $\{transection, \ normalizer\}$ specifies the symmetric non-degenerate second order tensor up to multiplication to one of the fourth roots of unit.

Notice that the reversing of sign of the normalizer does not affect the family of tensors ('metrics') mentioned in proposition. This is useful to take in mind if one deals with non-orientable manifold where continuous field of normalizer (nonzero 4-form) might not exist. The existence of 'field' of nonzero 4-form defined up to sign does not require orientability of manifold and it is sufficient for construction of the 'field' of metric fixed up to multiplication to $1^{1/4}$.

More definitely the relation of transection and metric requires introduction of the following notion.

> **Definition 2.7**  Let us consider the set of collections $\{transection, \ the \ ordering \ of$ *lobes, normalizer, transection orientation*$\}$. We will call two such objects *equivalent* if they are either identical or
>    (i) both transections coincides *and*
>    (ii) the ordering of lobes are opposite *and*
>    (iii) normalizer from one collection equals minus normalizer from another one *and*
>    (iv) if $S_{\dot{A}\dot{B}}$ is the oriented $S$-basis of dotted lobe from the first collection then the

triad of 2-forms

$$\tilde{S}_0 = S_{\dot{0}}, \quad \tilde{S}_1 = S_{\dot{1}}, \quad \tilde{S}_2 = S_{\dot{2}},$$

is the oriented $S$-basis of the undotted lobe from the second collection.

Of course the last condition is meaningful only when all the previous ones are fulfilled.

≀ **Definition 2.8**  The class of objects equivalent in accordance with definition 2.7 constitutes *normalized ordered oriented transection*.

The following basic theorem holds.

≀ **Theorem 2.9**  Every normalized ordered oriented transection determines unique metric tensor and *vice versa*.

## 3. Closed characteristic of transection lobe

In accordance with the initial definition of a transection lobe (see definition 1.1 and discussion below it) it is specified in terms of relation to another lobe which, due to proposition 1.4, cannot involve any arbitrariness and, in turn, is determined by the same relation to the first one. It is desirable to break this loop of definitions and to develop a closed description of a lobe in terms of its algebraic properties as a subspace of $\Lambda^2$. Such a result would enable one to get a better insight to the sense of the transection notion providing it by an indication of logical completeness.

To begin with, let us introduce the following useful auxiliary notion.

≀ **Definition 3.1**  A subspace $^?\Lambda^2$ of $\Lambda^2$ will be called *complete* if for every simple nonzero 2-form $\alpha \in {}^?\Lambda^2$ there exist a *simple* 2-form $\beta \in {}^?\Lambda^2$ such that $\alpha \wedge \beta \neq 0$.

In accordance with remark 1.6 the completeness notion is not vacuous for more than 1-dimensional subspaces.

Preliminary description of a lobe is provided by the following criterion.

≀ **Proposition 3.2**  3-dimensional subspace $^?\Lambda^2$ of the space of all 2-forms $\Lambda^2$ is a lobe of some transection if and only if it is complete.

It holds for both lobes of course.

The proof (see Appendix) is based on the following simple remark.

≀ **Lemma 3.3**  For every nonzero 2-form $\alpha$ belonging to (say, undotted) lobe of transection there exists an $S$-basis (normalized if desirable) of the lobe such that $\alpha = S_0$ for simple $\alpha$ and $\alpha = kS_1$ for non-simple $\alpha$ ($k$ is some nonzero number).

For the dotted lobe one has a similar statement of course.

The lobe description provided by the proposition 3.2 becomes more transparent if one takes in account the following fact:

≀ **Proposition 3.4**  Any two nonzero simple elements of the lobe of some transection are not wedge orthogonal.

It is implied, in particular, by the second separability criterion:

≀ **Theorem 3.5**  3-dimensional subspace $^?\Lambda^2$ of $\Lambda^2$ is a lobe of some transection if and only if its every 2-dimensional subspace contains a non-simple 2-form.

**Remark 3.6** Due to remark 1.6 and lemma 1.7 every 2-dimensional subspace of a lobe is spanned by a pair of simple and non-simple elements. The vanishing or non-vanishing of their wedge product is the invariant characteristic of subspace. If it vanishes the simple element is unique (up to a scalar factor) otherwise there are two 'simple directions' in subspace.

Theorem 3.6 forbids a lobe to possess of more than 1-dimensional subspaces consisting of simple elements only. It is convenient to introduce the following terminology:

≀ **Definition 3.7** More than 1-dimensional linear subspace of $\Lambda^2$ will be called a *simplest* one if the wedge product of any two its elements (including coinciding ones) vanishes.

In particular a simplest subspace consists of simple elements only. The reversing statement holds as well: if subspace consists of the simple elements only it is simplest. Simplest subspaces are of course not complete. A useful example of the simplest 3-dimensional subspace of $\Lambda^2$ is one spanned by the elements $\Theta^1 \wedge \Theta^2, \Theta^2 \wedge \Theta^3, \Theta^3 \wedge \Theta^1$.

Now the theorem 3.5 can be recast as follows:

≀ **Theorem 3.5'** 3-dimensional subspace $^?\Lambda^2$ of $\Lambda^2$ is a lobe of some transection if and only if it admits no simplest subspaces.

It turns out however that *a priori* restriction to the dimension of a 'candidate' to transection lobes can be dropped out due to the following

≀ **Lemma 3.8** Any subspace of $\Lambda^2$ whose dimension exceeds 3 contains a simplest subspace.

It implies the following main criterion:

≀ **Theorem 3.9** The subspace $^?\Lambda^2$ of $\Lambda^2$ is a lobe of some transection if there exists its complement $^!\Lambda^2$, $\Lambda^2 = {^?\Lambda^2} \oplus {^!\Lambda^2}$, such that neither $^?\Lambda^2$ nor $^!\Lambda^2$ contain simplest subspaces.

(We use here these somewhat exotic notations in order to emphasize that $^?\Lambda^2$ may be interpreted either as $^+\Lambda^2$ or $^-\Lambda^2$ and then $^!\Lambda^2$ will be $^-\Lambda^2$ or $^+\Lambda^2$ respectively).

The requirements of the latter theorem can be considered as *another definition* of transection alternative to initial definition 1.1. It is important that here we have no *a priori* restrictions on the dimensions of lobes.

≀ **Definition 3.10** The transection of the space $\Lambda^2$ of 2-forms over $\mathbb{C}$ is its decomposition into a direct sum of two subspaces such that neither of them possesses simplest subspaces.

The conditions of applicability of the theorem 3.5 are obviously stable with respect to slight "motions" of the subspace $^?\Lambda^2$. In more precise terms this means that the set of all (3-dimensional) subspaces of $\Lambda^2$ which may be considered as lobes of some transections (*i.e.* in accordance with the theorem 1.21 produce some conformal metrics) is open in the space of all 3-complex-dimensional subspaces of $\Lambda^2$, the *Grassmann manifold* $\mathbb{C}G_{6,3}$. Moreover, such set is dense in $\mathbb{C}G_{6,3}$ as well. We have the next basic theorem.

**Theorem 3.11** The set of all complex conformal metrics (over a point) is in 1-to-2 correspondence with the *open dense* subset of the Grassmann manifold of 3-dimensional subspaces of 6-dimensional complex vector space.

This map is a twofold mapping because every two subspaces of $\Lambda^2$ that are the opposite lobes of some transection are distinct as points of Grassmann manifold but determine the same conformal metric class.

## 4. 4-dimensional Riemannian structure and transection bundle

In principle one may suggest a number of different ways of representation of the space-time metric structure.

The basic one is the using of a symmetric non-degenerated second rank tensor field. But the conformal metric can be also exhaustively described by means of the specifying of the set of null cones formed by null geodesics emanating from a point which in turn runs through the whole space-time manifold (see for example [47]). If additionally the volume element (some non-zero 4-form) is specified a proper alternative description of the metric structure arises.

The approach based on transections yields another 'independent' description of the Riemannian structures although it suits to the case of dimension 4 alone.

Indeed, it is natural to consider all the constructions introduced above to be based on the exterior cotangent algebra over a point of some 4-dimensional manifold (real or complex holomorphic) which will be called a space-time for brevity. If one 'attaches' a transection to every point and makes this in a special 'continuous and smooth' manner the description of the *conformal* metric structure alternative to canonical one arises. In a similar way a 'continuous and smooth' field of normalized ordered oriented transections yields a proper description of the ordinary metric properties of space-time (It is evident that for complex holomorphic manifold the metric structure will be complex holomorphic, meanwhile in the 'real' case most natural way leads just to Lorentzian metrics as we shall see below). Every metric can be realized in such a way, at least locally.

Further, if one need to define a notion of 'global field' of some object in may achieve this by means of explicit definitions of 'local fields' of this object in coordinate neighborhoods and, additionally, the rules connected such local fields in neighborhood intersections. Another and maybe more plausible way which yields a rigorous base for a former one as well is to introduce some fiber bundle over underlying manifold and then the fields may be realized as the sections of the bundle in according with general theory [48,49]. The receipt of construction of such a fiber bundles assumes specification of the group which consists of transformations mapping arbitrarily chosen object to another arbitrarily chosen one (structure group) and another group that retains every object unaffected.

We shall outline in this section some primitives for such *transection bundle* handling. The word 'object' here means normalized ordered oriented transection, the space where the groups acts is $\Lambda^2$.

Let a (pseudo- or complex-)Riemannian space-time be covered by a family of connected opened sets (coordinate neighborhoods) such that every open set is parallelizable and there are smooth (holomorphic on complex manifold) covector fields over it (a tetrad of 1-forms) that yields the metric tensor fields by equation (1.11) (or, equivalently, (1.12)). Then the formulae (1.3) yield the smooth fields of undotted and dotted $S$-forms of conformable orientation satisfying identities (1.5-7) over the corresponding coordinate neighborhoods. Moreover, one has a normalized oriented ordered transection over every point of neighborhood. Their collection may be called a local transection field. Local normalizer field can be determined by the formula $\omega = (3i)^{-1} S^{AB} S_{AB}$.

Over a point common for two coordinates neighborhoods the corresponding tetrads of orthonormal 1-forms are connected by the $O(4, \mathbb{C})$ transformations and this transformation smoothly (or holomorphically) depends on a point. The group acting on $S$-forms is described as follows. Using lemma 1.15 one can show that lifting to the level of $S$-bases the orthonormal relation of the bases of 1-forms implies that $S$-forms are connected by composition of

(i) $SO(3, \mathbb{C})$-equivalent rotations independent for dotted and undotted $S$-bases that 'keep the lobes unchanged' and

(ii) mutual interchange of undotted and dotted $S$-bases

(that changes lobes interchanging them but retains the transection itself; it can be connected with reflection of some element of tetrad, *i.e.* with non-proper transformation from $O(4, \mathbb{C})$).

These transformations of $S$-forms are exactly those which do not affect the metric tensor connected with chosen fields of $S$-bases. We may specify the corresponding group $I$ acting in $\Lambda^2$ and keeping the metric connected with $S$-bases as semidirect product of the following form

$$I = (SO(3, \mathbb{C}) \times SO(3, \mathbb{C})) \times_j \mathbb{Z}_2.$$

The factors in parentheses are the undotted and dotted rotations groups, the only nontrivial element of $\mathbb{Z}_2$ acts on factors by means of the map $j$ which simply interchanges them.

Let us notice now that eqs. (1.5-7) being necessary and sufficient restrictions that guarantee $S$-bases to be connected with some transection are at the same time equivalent to requirement for the following collection of 2-forms

$$
\begin{aligned}
\Sigma_1 &= S_{\dot{0}} + S_{\dot{2}}, & \Sigma_2 &= iS_{\dot{0}} - iS_{\dot{2}}, & \Sigma_3 &= 2S_{\dot{1}}, \\
\Sigma_4 &= iS_{\dot{0}} + iS_{\dot{2}}, & \Sigma_5 &= S_{\dot{2}} - S_{\dot{0}}, & \Sigma_6 &= 2iS_{\dot{1}}
\end{aligned}
\tag{4.1}
$$

to be weakly orthonormal with respect to wedge product. This means that only $\Sigma_j \wedge \Sigma_j$ (without summation over $j$) are nonzero and the values of these 6 products are the same for all $j$.

We see therefore that any two collections of $S$-bases that correspond to some transections (which may be different) are connected by some transformation from 6-dimensional conformal group, a direct product $\mathbb{C}^* \times O(6, \mathbb{C})$ where $\mathbb{C}^*$ is the multiplicative group of nonzero complex numbers.

It is insufficient however for $S$-forms to obey eqs. (1.5-7) in order to determine the fitted $S$-bases. It is clear that exactly one from two weakly orthonormal $\Sigma$-bases that differ only by orientation (in the sense of linear spaces theory) yields fitted $S$-bases because the change of orientation of $\Sigma$-bases (4.1) can be realized as the reversing of signs of their first three elements that in turns implies conformal reflection of undotted lobe. Thus if one constructs the 6-basis $\tilde{\Sigma}_j$ by means of $O(6, \mathbb{C})$ transformation from a given 6-basis $\Sigma_j$ which is implied by some fitted $S$-bases then exactly one of two collections $\tilde{\Sigma}_j$ and $\hat{\Sigma}_j = \{-\tilde{\Sigma}_a, \tilde{\Sigma}_{a+3}\}$, $a = 1, 2, 3$ will yield the fitted $S$-bases (and therefore some metric). But the transformation $\tilde{\Sigma}_j \rightarrow \hat{\Sigma}_j$ is a non-proper $O(6, \mathbb{C})$-rotation in fact. Then the connectedness reasons suggest that the relevant orthonormal transformations *mapping transection to transection* are exactly proper ones. Thus one must narrow down the above mentioned group to special conformal group $CSO(6, \mathbb{C}) = \mathbb{C}^* \times SO(6, \mathbb{C})$.

Thus the existence of *global* field of fitted $S$-bases implies the existence of a section of a principal bundle with the structure group $CSO(6, \mathbb{C})$ (and then this bundle will be trivial). But such a requirement is too strong for our purposes. We need to define the global field of normalized oriented ordered transections, not $S$-bases themselves. The former correspond to the fiber bundle associated with mentioned above whose fiber $T$ is a factor space with respect to group $I$ which changes $S$-bases but does not affect transection:

$$T = CSO(6, \mathbb{C})/I = \mathbb{C}^* \times SO(6, \mathbb{C})/\{(SO(3, \mathbb{C}) \times SO(3, \mathbb{C})) \times_j \mathbb{Z}_2\}.$$

The fields of $S$-bases exist over coordinate neighborhoods but in their intersections corresponding $S$-bases do not coincide in general being connected by transformations from the group $I$.

One can easily show that the group $CSO(6, \mathbb{C})$ acts on the space $T$ effectively since $I$ does not involve any its normal subgroup [48].

    **Definition 4.1** We shall call the bundle over space-time manifold with the structure group $CSO(6, \mathbb{C})$ and typical fiber $T$ *the transection bundle*.

One has the following non-local version of the theorem 2.9:

    **Theorem 4.2** The complex metric structure on the 4-dimensional manifold (real or complex holomorphic) can be specified by a section of the transection bundle over this manifold.

As usually, the sections of the transection bundle are interpreted as fields of normalized ordered oriented transections.

It is worth noting that existence of a section of the transection bundle does not imply generally speaking the existence of a field of non-zero 4-form (orientation in case of real manifold). The normalizer yielded by the section is a smooth field only over the coordinate neighborhood but it may reverse the sign during transition to another neighborhood in their intersection.

## 5. Curvature endomorphism

Besides the metric tensor itself which is encoded in our formalism by the oriented ordered normalized transection the description of metric space includes also the tensor object of a compatible importance, the curvature. It may be represented in a number of alternative forms and here we shall see how the curvature notion can be translated to the transection language.

It is well known that local fields of triads of anti-self-dual and self-dual 2-forms that are defined in terms of null tetrads by the formulae (1.3) and obey the eqs. (1.5-7) (*i.e.* $S$-bases in our terminology) can be used for effective and exhaustive description of the Levi-Civita connection associated with the metric (1.11). With connections in hands, the Riemannian curvature can be introduced in a natural way as well (corresponding equations will be given below). Such a description, however, inevitably involves additional specific arbitrariness which is usually called gauge freedom and is due to the fact that metric does not specify $S$-forms uniquely. Nevertheless it turns out that if one transforms the curvature description mentioned above to the language of transection fields then the gauge arbitrariness disappears and a rather natural and transparent independent realization of the curvature of Riemannian space comes of. The 'alternative metric description' can therefore be lifted to a level of curvature.

To clarify the approach based on the use of transection fields let us remind at first how a curvature (and connection) is described in terms of $S$-bases.

The *undotted connection 1-forms* $\Gamma_{AB} = \Gamma_{BA}$ (3 different components) obeys *the first structure equations*

$$dS_{AB} - 2\,\Gamma^C_{(A} \wedge S_{B)C} = 0 \tag{5.1}$$

and are determined by them in a unique way *provided* the eqs. (1.5-7) are satisfied. The inversely dotted version of eq. (5.1) yields dotted connection forms $\Gamma_{\dot{A}\dot{B}} = \Gamma_{\dot{B}\dot{A}}$. The collection $\{\Gamma_{AB}, \Gamma_{\dot{A}\dot{B}}\}$ involves all the information on the Levi-Civita connection associated with metric (1.11).

Next, *the second structure equations*

$$\Omega^A_B = d\,\Gamma^A_B + \Gamma^A_C \wedge \Gamma^C_B \tag{5.2}$$

determine *undotted curvature 2-forms* $\Omega_{AB} = \Omega_{BA}$. The inversely dotted equations hold as well and determine dotted curvature 2-forms $\Omega_{\dot{A}\dot{B}} = \Omega_{\dot{B}\dot{A}}$. The collection $\{\Omega_{AB}, \Omega_{\dot{A}\dot{B}}\}$ describes comprehensively the corresponding Riemannian curvature.

The undotted and dotted curvature forms admit the expansions with respect to the $S$-bases of the following form:

$$-\Omega^A_B = C^{AC}_{BD}\, S^D_C + C^{A\dot{C}}_{B\dot{D}}\, S^{\dot{D}}_{\dot{C}} - \tfrac{1}{12}R\, S^A_B, \tag{5.3a}$$

$$-\Omega^{\dot{A}}_{\dot{B}} = C^{\dot{A}\dot{C}}_{\dot{B}\dot{D}}\, S^{\dot{D}}_{\dot{C}} + C^{C\dot{A}}_{D\dot{B}}\, S^D_C - \tfrac{1}{12}R\, S^{\dot{A}}_{\dot{B}}, \tag{5.3b}$$

Here $C_{ABCD} = C_{(ABCD)}$ and $C_{\dot{A}\dot{B}\dot{C}\dot{D}} = C_{(\dot{A}\dot{B}\dot{C}\dot{D})}$ are the spinor components of the conformal curvature (constituting *undotted and dotted Weyl spinors*), $C_{AB\dot{C}\dot{D}} = C_{(AB)(\dot{C}\dot{D})}$ are the spinor components of the traceless part of the Ricci tensor (*Ricci spinor* respectively) and $R$ is the scalar curvature.

We define these spinor objects by the expansions (5.3) but it is also possible to connect them with the standard definitions via the null tetrad which is implied by eqs. (1.3). Then the curvature spinors component turn out to be the $\mathbb{C}$-linear combinations of various contractions of Riemann tensor (and its dual) with elements of tetrad. There are no reasons however for us to involve this primordial interpretation and eq. (5.3) will be considered here as basic definitions of irreducible curvature spinors.

It is worth noting that the expansions (5.3) are not of the most general possible form. For example, they does not involve such terms as $C^C_{(A}S_{B)C}$ and $C^{\dot C}_{(\dot A}S_{\dot B)\dot C}$ for some symmetric $C_{AB}$, $C_{\dot A\dot B}$ respectively (assuming that all free indices in (5.3) are lowered). Additionally, the same spinor $C_{AB\dot C\dot D}$ and scalar $R$ enter different eqs. (5.3a) and (5.3b) that also would not be the case for generic expansions. These features reflect in fact the well known characteristic symmetries of Riemann tensor and moreover they are equivalent to them.

In order to define the curvature globally one need not assume that the $S$-form and, accordingly, connection and curvature forms are defined on the whole underlying manifold. It is sufficient to specify them over every coordinate neighborhood that constitute the atlas covering the manifold. On the chart intersection they have to be connected by certain transformations of course. For $S$-forms corresponding transformations are undotted and dotted rotations possibly accompanied by the transformation described in the item (iv) of definition 2.7. These transformations yield certain transformations of the connection and curvature forms implied by the structure equations. Thus an additional index enumerating charts should be attached strictly speaking to all objects involving in the above formulae.

As to the transformation rules for the curvature forms the following fact is well known:

**Theorem 5.1** When the field of $S$-bases is undergone a rotation (smoothly dependent on a point) the curvature 2-forms are transformed exactly in the same way. Formally

$$\Omega[gS] = g\Omega[S].$$

Transformations of connection forms can be easily obtained as well but we shall not need them here.

The formulae (5.2) may be considered as local maps of the *set of local fields of $S$-bases* to $\Lambda^2$ (considered here as the space of fields of 2-forms over space-time manifold or some coordinate neighborhood) but really there exist a much stronger relation. To see this, let us introduce the following

**Definition 5.2** Let a local field of normalized ordered oriented transections be given. Let us define for every local field $\alpha$ of elements of undotted lobe induced by transection a local 'field' of 2-form $\Omega^+(\alpha)$ specifying its value in every point $p$ of neighborhood as follows:

- if $\alpha_{|p} = 0$ then $\Omega^+(\alpha)_{|p} = 0$;
- if $\alpha_{|p} \neq 0$ but $(\alpha \wedge \alpha)_{|p} = 0$ then there is a normalized oriented undotted $S$-basis such that $\alpha_{|p} = S_{0|p}$. $S$-basis $S_{AB}$ implies curvature 2-forms $\Omega_{AB}$ and we define $\Omega^+(\alpha)_{|p} = \Omega_{0|p}$;
- in the last case $(\alpha \wedge \alpha)_{|p} \neq 0$ there exists a normalized oriented undotted $S$-basis such that $\alpha_{|p} = kS_{1|p}$ for some nonzero function $k$, at least in some vicinity of $p$. $S$-basis $S_{AB}$ implies unique curvature 2-forms $\Omega_{AB}$ and we define $\Omega^+(\alpha)_{|p} = k\Omega_{1|p}$.

The map $\Omega^+ : {}^+\Lambda^2 \to \Lambda^2$ may be called *undotted curvature map*. The *dotted curvature map* $\Omega^- : {}^-\Lambda^2 \to \Lambda^2$ can be defined in a similar way with the only difference that the dotted $S$-bases must be used and they must be considered to be oriented if they are fitted with oriented undotted $S$-bases. Besides, the normalizing is understood in accordance with eq. (2.1$b$).

Due to unique decomposition $\Omega = \Omega^+ \oplus \Omega^-$ implied by transection one may define now the *curvature map*

$$\Omega = \Omega^+ \oplus \Omega^- : \Lambda^2 = {}^+\Lambda^2 \oplus {}^-\Lambda^2 \to \Lambda^2$$

which depends of course on transection entering the above definition.

**Theorem 5.3** The curvature map is correctly and uniquely defined and is an endomorphism. It is determined by the field of normalized ordered oriented transections and exhaustively characterizes the Riemannian curvature as it is usually defined.

Indeed, the definition 5.2 provides us with a *constructive* definition of the curvature map but does not guarantee that the result will be unique, continuous and smooth because some arbitrariness due to possible different choices of $S$-bases involved in it. Additionally different items of definition have to be matched continuously and smoothly which also needs a separate proof. Theorem 5.2 yields such a necessary proof of the correctness of definition 5.2.

The following remark is rather essential.

**Remark 5.4** The fact that the curvature tensor for Riemann space is in fact a field of endomorphisms on the bundle of second order antisymmetric tensors (2-forms) is of course well known. These endomorphisms can be restricted to the spaces of field of all anti-self-dual and self-dual 2-forms (that is equivalent to decomposition $\Lambda^2 = {}^+\Lambda^2 \oplus {}^-\Lambda^2$ yielding the maps formally coinciding with our $\Omega^+, \Omega^-$). The corresponding formulae will be of the form

$$Riem : S_{\alpha\beta} \to S_{\mu\nu} \mathcal{R}_{\alpha\beta}{}^{\mu\nu}$$

(the Greek letters are the tensorial indices here). In our approach a similar equation is observed as well but a rather different sense is to be attributed to it. Indeed, in the Riemann-Christoffel interpretation, the curvature $\mathcal{R}_{\alpha\beta}{}^{\mu\nu}$ is the derivative of the metric tensor, it is defined (and determined) in terms of its components without any relation to $S_{\alpha\beta}$ which is in turn arbitrary antisymmetric tensor having no relation to the metric there. As to our approach, we determine in fact the whole 'aggregate' $S_{\mu\nu} \mathcal{R}_{\alpha\beta}{}^{\mu\nu}$ (in the way specified by definition 5.2 in fact) and this is made *without any explicit reference to the metric tensor*. Of course, afterwards one may separate the components $\mathcal{R}_{\alpha\beta}{}^{\mu\nu}$ and they will coincide with ones calculated in a classical way but in any case they are not directly connected with the metric (of course if it is not regarded in a generalized way as 'property of metricity') as far as one speaks about their origin. In our case the curvature endomorphism is determined by [normalized ordered oriented] *transection* itself and could be named a *transection curvature* as well.

Although the values the of curvature map are really calculated by means of the formulae (5.1,2) the map is of more fundamental nature then simply their repetition. Meanwhile the eqs. (5.1,2) may be considered at best as a map (second order differential operator in fact) from the set of $S$-bases to $\Lambda^2$, the curvature endomorphism is the endomorphism of $\Lambda^2$ *itself* and does *not* depend on any definite $S$-forms. It is determined by the transection and exhaustively characterizes, in particular, the curvature of metric associated with transection. Together with the notion of transection itself the curvature map can be used as a second basic object for the 'alternative' description of the metric structure.

**Remark 5.5** Unfortunately, the curvature endomorphism depends on the 'initial' choice of the ordering and orientation of transection field but these different choices yields simply multiplication to a constant from a finite set.

It is worth noting here that two 'halves' $\Omega^+, \Omega^-$ of the curvature map are not independent. The basic cause of this fact is a unique fixation of the second lobe of transection when the first one is specified (proposition 1.4). Since either curvature half-map $\Omega^+$ or $\Omega^-$ are determined by the corresponding fields of lobes we may conclude that the *both* half-maps are implied by the *one sort* lobes distribution. It is reasonable to conjecture further that the lobe is determined by the corresponding curvature half-map since such a link can be formally realized as a system completely integrable equations (5.1,2) (together with the corresponding 'undotted Bianchi identities' which are not considered here) although this dependence is not 1-to-1. Then the first lobe implies the field of opposite one and finally the field of the opposite curvature half-map. In this sense the opposite halve of the curvature map is connected with the first one and in this sense they are mutually dependent. In any case, a knowledge of the only one (undotted or dotted) substructures enables one to restore the whole transection (and metric) structure.

Due to expansions (5.3) and taking into account an obvious fact that every individual component $\Omega_{AB}$ (or $\Omega_{\dot{A}\dot{B}}$) is an image of some element of $^+\Lambda^2$ ($^-\Lambda^2$ respectively) under the curvature endomorphism we have the following ɫ **Proposition 5.6** The normalized ordered oriented transection field specifies a vacuum solution of the Einstein equations if and only if the half $\Omega^+$ (or $\Omega^-$) of the curvature endomorphism is the endomorphism of the lobe $^+\Lambda^2$ (respectively $^-\Lambda^2$) itself. Then the opposite half of curvature map possesses the same property.

**Remark 5.7** The locally flat space-time corresponds to trivial curvature endomorphism. But although the halves of the curvature map are in a certain sense dependent it is well known that the vanishing of one of them does not imply the vanishing of another. In such a case this second non-vanishing half however will be necessary a lobe endomorphism (Ricci flat and conformally semi-flat spaces or heavens[12,13]).

**Remark 5.8** The case of Einstein equations with $\lambda$-term is assumed in the proposition 5.6. The vanishing of $\lambda$-term may be ensured by additional requirement for one of the halves of curvature endomorphism to be traceless (and then another half will be traceless as well).

In conclusion of section we shall outline how the curvature can be associated with field of transection when normalizer is *not* specified. Then transection field determines a *conformal class* of metrics rather than a single metric. Hence it is natural to associate with field of non-normalized transection the *conformal* curvature. For, assume that some $S$-basis of undotted lobe is chosen (at least locally). The conformal curvature can be defined as collection of 2-forms $\Omega_{AB}^{(W)} = \Omega_{BA}^{(W)}$ accumulating only Weyl spinor dependent part of the expansion (5.3a), *i.e.*

$$\Omega_{AB}^{(W)} = C_{AB}^{KL} S_{KL}.$$

Equivalent definition is provided by the following set of equations:

$$\Omega_{AB}^{(W)} \wedge S_{KL} = \Omega_{(AB} \wedge S_{KL)}, \quad \Omega_{AB}^{(W)} \wedge S_{\dot{C}\dot{D}} = 0$$

which determines $\Omega_{AB}^{(W)}$ uniquely.

The transformation properties of conformal curvature 2-form apparently coincide with those of ordinary curvature forms $\Omega_{AB}$ that are indicated by theorem 5.1 and therefore in case of normalized ordered oriented transection conformal curvature map can be constructed in a close analogy with definition 5.2. It will be an endomorphism of every lobe when restricted to it. We shall not discuss this application of conformal curvature but revert to case of non-normalized transection.

It is easy to see that $\Omega_{AB}^{(W)}$ is not affected by any rescaling of $S_{AB}$. Indeed, in accordance with eq. (5.1) if the connection $\Gamma_{AB}$ is associated with $S$-basis $S_{AB}$ then connection $\tilde{\Gamma}_{AB}$ associated with $\tilde{S}_{AB} = e^\phi S_{AB}$ ($\phi$ is arbitrary scalar) will be the following

$$\tilde{\Gamma}_{AB} = \Gamma_{AB} + \tfrac{1}{2}\phi_{(A}{}^{\dot{L}}\theta_{B)\dot{L}}$$

where $\phi^{A\dot{B}}$ are defined by expansion $d\phi = \phi^{A\dot{B}} \theta_{A\dot{B}}$. Then eqs. (5.2) imply that the first term of expansion (5.3a) remains unaffected by this transformation (that express the well known conformal invariance property of the Weyl curvature).

One may proceed analogously with definition 5.2 and define the conformal map $^{+}\Omega^{(W)}$ as follows. If 2-form $\alpha \in {}^{+}\Lambda^2$ is simple we chose any $S$-basis $S_{AB}$ obeying condition $S_0 = \alpha$, determine conformal curvature 2-forms $\Omega_{AB}^{(W)}$ associated with this $S$-basis and define $^{+}\Omega^{(W)}(\alpha) = \Omega_0^{(W)}$. Similarly if $\alpha$ is non-simple $S$-basis is chosen obeying the restriction $S_1 = \alpha$ and we define $^{+}\Omega^{(W)}(\alpha) = \Omega_1^{(W)}$. For vanishing $\alpha$ $^{+}\Omega^{(W)}(\alpha)$ is not defined. (The 'dotted' case does not require a separate explanation of course.)

There is a problem with this definition however. One may show that in case of *simple* $\alpha$ we obtain in fact not a single value but the 'conformal class' of 2-forms, *i.e.* $^{+}\Omega^{(W)}$ is defined up to arbitrary nonzero complex factor. On the other hand $\alpha$'s that differ by a factor will be sent by $^{+}\Omega^{(W)}$-map to the same 'conformal class' of 2-forms. Thus it is natural to assume that $^{+}\Omega^{(W)}$ acts on the *projective* lobe space (*i.e.* the set of classes of nonzero $^{\pm}\Lambda^2$ elements when proportional elements are considered to be equivalent) and maps it into itself.

When we consider *non-simple* arguments the situation is different. Now $^{+}\Omega^{(W)}(\alpha)$ is defined *up to sign* and elements distinguished by sign have the same image. Here the conformal curvature map acts obviously on $^{+}\Lambda^2$ quotient by $\mathbb{Z}_2$.

We obtain the pair of maps with apparently encode all the information on the conformal curvature associated with conformal metric which in turn is determined by transection. It is not quite clear however what would be the best way to incorporate these maps to a single one or, to be more exact, how properly describe the set on which it acts. Probably it is reasonable to consider the set of simple $S$-forms as boundary of the set of non-simple ones and represent the quotient sets mentioned above as some projective space with boundary. This problem remains still open however.

## 6. Reality conditions

Up to now we have dealt with linear spaces over the field of complex numbers (even if underlying manifold is real). It was an essential restriction because most relations stated above fail when one passes to case of real spaces. Indeed, the algebraic proofs often require to solve quadratic equations which in real case may have no solutions.

In general the metric implied by transection (together with associated $S$-forms, tetrads, *etc.*) is complex valued even if it is evaluated on the space of real vectors. But sometimes it might turn out that the conformal class of metrics associated with transection contains a real one and correspondingly normalized ordered oriented transection yields a real metric. The goal of this section is to describe the necessary and sufficient conditions implying such a situation which is of the main physical interest.

One may consider the complex holomorphic or real 4-manifolds. In former case the $(n, 0)$ differential forms bundles are used and the metric (1.11) will be complex holomorphic. It latter case *complexified* bundles are used instead. Then the metric (1.11) will be generally complex as well (as bilinear function on a tensorial square $T_{\mathbb{R}} \otimes T_{\mathbb{R}}$ of the real tangent space $T_{\mathbb{R}}$) but a case of real metric is also possible.

The complexification of cotangent and higher degree Grassmann bundles naturally arises in the intermediate situation when space-time is considered as a special sort submanifold (real slice, see for example [50]) embedded into some 4-complex-dimensional holomorphic manifold endowed with holomorphic (not Kählerian) metric. The importance of complex holomorphic Riemannian spaces is in part due to the fact that in many cases real Einstein spaces, especially vacua and electrovacs, may be interpreted, at least locally, as real slices of holomorphic solutions of (suitably modified) 'complex' Einstein equations [50]. Indeed, most of the vacuum and electrovac metrics involves only real analytical functions of coordinates and parameters and it is sufficient to allow latter to be complex to obtain a local complex holomorphic Einstein space-time.

Conversely when each of the complex coordinates is restricted to some 1-dimensional submanifold of the complex plane a real metric may sometimes result, the Einstein equations remaining valid. Furthermore if we begin with a real metric a different metric may be obtained and relations between different classes of space-times may be established. Another useful trick is to divide the problem of the searching for solutions of Einstein equations into two steps: at first, to construct the complex solution and, at second, to try to find its real slices.

The method of restriction of the holomorphic space to a real slice naturally puts into consideration the complexification of the cotangent bundle and its derivatives rather that real ones as the pull-back of the corresponding holomorphic bundles under the embedding map. A 'restriction' (pull-back in fact) of the transection bundle is done in a natural way and transection fields on the holomorphic space-time induce transection fields on a real slice.

Let us consider a real 4-dimensional manifold (possibly a real slice of some holomorphic manifold) and a foliation $^+\Lambda^2$ of complex valued 2-forms on it. Let the field of transections of $^+\Lambda^2$ be given.

The following simple statement

↲ **Lemma 6.1**  If $\alpha \wedge \alpha = 0 = \beta \wedge \beta$ for some nonzero $\alpha \in {}^+\Lambda^2$, $\beta \in {}^-\Lambda^2$ then there exist local fitted $S$-bases such that $\alpha = S_0$, $\beta = S_{\dot{0}}$.

is useful in proof of the theorem which clarifies the condition ensuring existence of real metric in the conformal class implied by transection field:

↲ **Theorem 6.2**  The local field of transections on a real 4-manifold satisfies condition

$$\overline{{}^+\Lambda^2} \subset {}^-\Lambda^2$$

(equivalent to $\overline{{}^-\Lambda^2} \subset {}^+\Lambda^2$) if and only if the corresponding conformal metric class contains a real metric of Lorentz signature.

Such a property of transection can be named *conjugating symmetry*. Of course, the inclusions can be replaced by equalities here.

**Remark 6.3**  The normalized oriented ordered transection field will determine a real metric if additionally normalizer is real.

↲ **Corollary 6.4**  Under the conditions of the theorem 6.2 there exist such $S$-bases of undotted and dotted lobe that

$$\overline{S_{AB}} = S_{\dot{A}\dot{B}} \qquad (0 \overset{c.c.}{\Longleftrightarrow} \dot{0}, \ 1 \overset{c.c.}{\Longleftrightarrow} \dot{1}); \tag{6.1}$$

moreover, they are fitted by such a basis $\theta_{A\dot{B}}$ of $\Lambda$ that

$$\overline{\theta_{A\dot{B}}} = \theta_{B\dot{A}} \qquad (0 \overset{c.c.}{\Longleftrightarrow} \dot{0}, \ 1 \overset{c.c.}{\Longleftrightarrow} \dot{1}). \tag{6.2}$$

("Hermitian" property of tetrad).

Of course, on case of conjugating symmetry $S$-bases and tetrads that do not obey eqs. (6.1,2) may be used as well.

↲ **Corollary 6.5**  Any nonzero element of either $^+\Lambda^2$ or $^-\Lambda^2$ may be neither real nor pure imaginary. Moreover, its real and imaginary parts are linearly independent.

Due to theorem 6.2 the theorem 3.5' can be complemented by the following its counterpart:

↲ **Theorem 6.6**  The set of local (complex-)conformal classes of real Lorentz signature metrics is in 1-to-2 correspondence with local fields $^?\Lambda^2$ of subspaces in complexified foliation $\Lambda^2$ of 2-forms such that

(i) $^?\Lambda^2$ does not contain simplest subspaces,

(ii) $^?\Lambda^2$ obeys equation $^?\Lambda^2 \wedge \overline{{}^?\Lambda^2} = 0$ and

(iii) $\Lambda^2 \subset {}^?\Lambda^2 + \overline{{}^?\Lambda^2}$.

The correspondence is 1-to-2 because subspaces $^?\Lambda^2$ and $\overline{{}^?\Lambda^2}$ determine the same conformal metric class.

It seems to exist no more simple 'reality' condition excluding maybe the case of 'real lobes' $^{\pm}\Lambda^2 = \overline{^{\pm}\Lambda^2}$ only (see also [32]). The collection of properties (i-iii) distinguishing Lorentz transections seems to be remarkably natural.

In the conjugation-symmetric case, it is possible to distinguish among all pairs of fitted $S$-bases ones obeying eqs. (6.1) (*conjugately fitted $S$-bases*). The group which preserves such a relation is constructed from the following subgroups:

$$g_0(\rho)\dot{g}_0(\overline{\rho}), \qquad g_1(\lambda)\dot{g}_1(\overline{\lambda}), \qquad g_2(\sigma)\dot{g}_2(\overline{\sigma}), \qquad g_\uparrow\dot{g}_\uparrow \qquad (6.3)$$

for arbitrary complex $\rho, \sigma$ and nonzero complex $\lambda$. The squared discrete rotations $g_\uparrow^2 = \dot{g}_\uparrow^2$ also preserve the conjugately fitting but they are degenerated to identical transformations (acting to $S$-forms, meanwhile they reverse the sign of tetrad elements) and may be omitted. It has been mentioned above that both $g_1, \dot{g}_1$ can be expressed in terms of other elementary rotations; moreover they can be connected with the group unit by continuous paths. The same is true of course for their product $g_1\dot{g}_1$ but additionally in the latter case the path cat be constructed from the 'real' rotations only. Thus $g_\uparrow\dot{g}_\uparrow$ belongs to the connected unit component. Other elementary real rotations in the list (6.3) obviously possess the same property. It is easy to show that all they can be interpreted as Lorentz transformations.

We see therefore that in conjugation-symmetric case the group of rotations acting on $S$-forms may be reduced to the proper orthochronous Lorentz group $SO^\uparrow(3,1)$.

The nontrivial $\mathbb{Z}_2$ element involved into the isotropy group $I$ must be taken into account as well. It can be interpreted either as a reflection of one of the spacetime directions of the space where $SO^\uparrow(3,1)$ acts or as complex conjugation of the parameters in transformations (6.3) ($g_\uparrow\dot{g}_\uparrow$ is not changed in such a case) when considered as automorphism of $SO^\uparrow(3,1)$. Then the semidirect product $SO^\uparrow(3,1) \times_{\text{c.c.}} \mathbb{Z}_2$ arises. It constitutes the isotropy group in the conjugately-symmetric case.

A simple analysis shows that structure group 'component' $SO(6,\mathbb{C})$ is reduced now to $SO(3,3)$ and conformal 'factor' $\mathbb{C}^*$ to $\mathbb{Z}_2 = \{1, -1\}$. Finally we see that the structure group in the conjugation-symmetric case is $\mathbb{Z}_2 \times SO(3,3)$. Consequently our local construction of conjugately-symmetric transection has the following global counterpart:

**Theorem 6.7** Lorenz metric corresponds to the field of conjugation-symmetric ordered oriented normalized transections which in turn may be realized as the section of fiber bundle with structure group $\mathbb{Z}_2 \times SO(3,3)$ and typical fiber $\mathbb{Z}_2 \times SO(3,3) / \left(SO^\uparrow(3,1) \times_{\text{c.c.}} \mathbb{Z}_2\right)$.

It is natural to name latter the *conjugately-symmetric transection bundle*.

## 7. Real transection without complexifications

We have see above that real Lorenz metrics corresponds to transections of *complexified* 2-forms space $\Lambda^2$. On the other hand complexification of a real space is equivalent to introduction of a twice-dimensional space plus a linear anti-involutive operator $J$ on it, which is called *complex structure*. In case of transections however there are no needs to employ twice-(*i.e.* 12-)dimensional space. Indeed, the transection is completely characterized by a single lobe, say $^+\Lambda^2$, and the lobe itself can be considered as a complexified space. Then its real 'source' is just the real 2-forms space $\Lambda^2_\mathbb{R}$. For, let us notice that it follows from $J^2 = -id$ that $\Re\alpha$ and $\Im\alpha$ are linearly independent for every $\alpha \in {}^+\Lambda^2$ and therefore $\Re\,{}^+\Lambda^2 + J\,\Im\,{}^+\Lambda^2 = \Lambda^2_\mathbb{R}$. Then the real transection turns out to be equivalent to the anti-involutive linear operator $J$ on $\Lambda^2_\mathbb{R}$ obeying additional restriction

$$J\alpha \wedge J\beta = -\alpha \wedge \beta \quad \forall \; \alpha, \beta \in \Lambda^2_\mathbb{R}. \tag{7.1}$$

(One may see that $J$ is nothing else but the dualizing operator for the metric connected with transection). Furthermore the lobes ${}^\pm\Lambda^2$ can be defined via $J$ as the sets (spaces in fact) of formal complex sums $\alpha \pm iJ\alpha$ respectively. Conversely if a transection is given and $\alpha \in {}^\pm\Lambda^2$ then we may define $J$ by equality $J(\Re\alpha) = \pm\Im\alpha$ and such an operator will be a complex structure satisfying (7.1).

We see therefore than one may banish the imaginary unit from the transection description of Lorentz metric introducing some operator $J : \Lambda^2_\mathbb{R} \to \Lambda^2_\mathbb{R}$ (cf. [40]). This receipt badly corresponds however with the principle that we have followed to above: to use *subspaces* (of 2-forms space $\Lambda^2$) rather than *operators* for metric structure representation. Fortunately, it turns out that one can proceed along these lines in the case of the description of transection in terms of real 2-forms only.

The problem is to describe an anti-involutive operator on $\Lambda^2_\mathbb{R}$ obeying (7.1) in terms of some subspaces (maybe subsets) of $\Lambda^2_\mathbb{R}$. This can be achieved and corresponding subset turns out to be a certain *projective quadric* in $\Lambda^2_\mathbb{R}$. Moreover, this quadric may be exhaustively characterized *as a subset* of $\Lambda^2_\mathbb{R}$ obeying a collection of properties that refer to the values of *binary exterior products* of its elements only and if these properties are observed the unique $J$-operator necessarily exists. One needs not even assume *a priori* for this set to be a submanifold in fact.

Unfortunately we cannot enter more details here and even state the main theorem because this would require too lengthy preliminaries, definitions requiring some proofs *etc.* This problem will be discussed in details elsewhere.

## 8. Summary

There is a lot of definitions, lemmas, theorems *etc.* in the above sections. It might be perhaps not so easy to separate important information from more technical details and that is why in this section we accumulate and describe in a brief 'qualitative' fashion most of essential statements.

We have seen that in case of four dimensions the Riemannian spaces, both complex holomorphic and real of Lorentz signature, admit alternative description which refers in no way to the field of metric tensor, the central object for ordinary introduction of such geometries. The notion of field of *transections* based on (holomorphic or complex valued respectively) foliation $\Lambda^2$ of 2-forms over space-time manifold may be used instead.

In each point, transection means simply decomposition of $\Lambda^2$ into a direct sum of two subspaces possessing a remarkable property to be *mutually wedge orthogonal* (definition 1.1). They are named here *lobes* of transection, undotted and dotted respectively, for brevity. The lobes of transection possesses another remarkable property that exhaustively distinguishes them as well: they do not admit *simplest subspaces*, *i.e.* such subspaces of more than 1 dimension on which binary operation of exterior multiplication yields only zeros (theorem 3.9). Moreover, it turns out that if $\Lambda^2$ is decomposed into a direct sum of such *complete* subspaces (see definition 3.1, proposition 3.2) then they are automatically 3-dimensional.

The set of subspaces of $\Lambda^2$ being a lobe of *some* transection turns out to be in 1-to-2 correspondence with an open dense subset in Grassmann manifold of complex-3-dimensional subspaces of complex-6-dimensional linear space $\Lambda^2$.

Being a 'homogeneous' object (direct sum of two linear subspace) transection specifies a conformal metric class rather than a single metric (and in turn is determined by latter in a unique way). It is worthwhile to mentioned here that if one deals with *real* space-time this conformal class will contain a real metric of Lorentz signature if and only if the complex conjugation acts on transection as non-trivial involution. In other words it sends every lobe of transection *into* (and *onto*) another lobe (theorem 6.2). Another way to describe such *conjugately symmetric* transections without complexification of the space of 2-forms is mentioned in section 8.

A reduction of conformal geometry to metric one is achieved by a certain normalizing procedure. *Normalizer* can be preliminary determined as nonzero 4-form, a volume element (real in case of real Lorentz geometry). More careful consideration shows however that one must take into account two additional relations that influence the metric implied by normalized transection. One of them is *orientation* of lobes. Basically, it is similar to their orientation as linear spaces and is 'reversed' when all three elements of the basis of lobe are multiplied to -1. In case of transection however there exists the canonical correspondence of the orientations of each lobe that allows one to introduce the *orientation of transection* itself (definition 2.5).

Another relation affecting the metric implied by transection is the *ordering of lobes*, *i.e.* a choice what lobe will be named 'undotted' and what 'dotted' one. This improper influence disappears however if the change of order of lobes is accompanied by the change of sign of normalizer (which as a result loses the rights to be considered as 'everywhere nonzero 4-form'; but another plausible consequence is that then the non-orientability of manifold is not longer an obstruction for normalizer to exist).

All these relations interfere and as a result we obtain the notion of *normalized ordered oriented transection* (see section 2).

The global realization of the above local constructions can be obtained in frames of fiber bundles language. They are the sections of a certain *transection bundle* (see section 4).

And the second after metric basic notion of Riemannian geometry, the notion of curvature, obtains its natural counterpart in a transection language. It is realized as *curvature endomorfism* (definition 5.2, theorem 5.3). In case of conformal geometry the *conformal curvature map* may be introduced.

We see that the most important attributes of Riemannian geometry can be expressed in terms of transection without any reference to metric tensor.

## 9. Conclusion

In frames of alternative approach, the metric geometry has two levels of description connected by equivalence relation. On the first level, Riemannian space in every chosen chart covering space-time manifold is described by the smooth field of special 3-dimensional oriented subspaces of the 6-dimensional space $\Lambda^2$ of complexified 2-forms together with everywhere nonzero 4-form, the normalizer. Such an aggregate realizes the local field of normalized ordered oriented transection. Its specification is sufficient for determination of local metric geometry.

In the intersections of charts the pairs {subspace,normalizer} must either coincide or *be equivalent* in a way described in definition 2.7. Then they constitute global field of normalized ordered oriented transections. In more rigorous terms such fields may be also defined as the sections of the fiber bundle over space-time manifold with, in complex holomorphic case, structure group $\mathbb{C}^* \times SO(6,\mathbb{C})/$ and typical fiber $\mathbb{C}^* \times SO(6,\mathbb{C})/\{(SO(3,\mathbb{C}) \times SO(3,\mathbb{C})) \times_j \mathbb{Z}_2\}$, and, in Lorentz case, structure group $\mathbb{Z}_2 \times SO(3,3)$ and typical fiber $\mathbb{Z}_2 \times SO(3,3)/(SO^\uparrow(3,1) \times_{c.c.} \mathbb{Z}_2)$.

Although these basic objects seems to be somewhat unusual there is a number of relations that allows one to translate to transection language most of notions existing in frames of standard metric description. Sometimes they may seems even a bit surprising. As an interesting example we have the following criterion distinguishing the type of covector:

(i) the real 1-form $h$ is timelike if and only if there exist no such simple nonzero elements $\alpha \in {}^+\Lambda^2$ that $h \wedge \Re\alpha = 0$;

(ii) it is spacelike if and only if there is a simple nonzero $\alpha \in {}^+\Lambda^2$ such that $h \wedge \Re\alpha = 0$ but there is no simple nonzero $\beta \in {}^+\Lambda^2$ such that $h \wedge \beta = 0$;

(ii) h is null if and only if there exist a simple nonzero $\beta \in {}^+\Lambda^2$ such that $h \wedge \beta = 0$.

Another important property admitting simple representation in terms of transections is geodetic one.

The second and more practical level of description of Riemannian structures in terms of transections (but without explicit use of this notion if desirable) is provided by introduction of nontrivial redundancy group, group of rotations. Then in spite of transection lobes the $S$-bases defined up to this group transformations are introduced. Such a formulation is well known and used in solving of many problems. The sets of $S$-forms are most often referred to as spaces of self- and anti-self-dual 2-forms. It is worth noting that introduction of $S$-bases does not require any additional requirements and locally is possible for every given transection field. Another relevant remark is that $S$-forms contains all the information on transection and when they are chosen the latter notion may be abandoned.

The $S$-forms can be used for construction of metric, but in principle this is not necessary for determination (and calculation) of such objects as connection and curvature forms. They may be obtained independently by means of eqs. (5.1-3) and inversely dotted ones. These equations are often used in practical computations. In case of real Lorentz geometry a useful 'complexification' is achieved and approximately twofold decreasing (at least formal but sometimes of decisive importance) of a number of field equations results.

A similar effect is observed in case of complex holomorphic geometry as well. There is no 'complexification' here of course but a simplification is achieved because it is sufficient to specify only half of the family of geometric objects (say, undotted ones) solving relevant equations and then all the geometry can be reconstructed together with the second half.

Thus the transection formalism combines both 'abstract' method of the global ge-ometry description in terms of sections of some fiber bundles and local representation of basic geometrical objects in the exterior forms language in a way rather convenient for various applications. Its power seems to be compatible with one of the standard geometry representation in terms of metric tensor.

It is difficult now to claim something definite on the comparative advantages and disadvantages of both descriptions of metricity. It seems clear however that transection approach is worthy of further investigations.

**Acknowledgments**. I am grateful to the *Graduertenkolleg* "Scientific computing" (*Köln – St. Augustin, Nordrhein-Westfalen*) for financial support and to the Institute for Theoretical Physics of the University of Cologne for hospitality.